\documentclass[useAMS,usenatbib]{mn2e}\usepackage{natbib}

\usepackage[latin1]{inputenc}
\usepackage{graphicx}
\usepackage{float}
\usepackage{color}
\usepackage{footnote}
\usepackage{amssymb}
\usepackage[figuresright]{rotating}
\usepackage{amsmath}
\usepackage{url}
\usepackage{rotating}
\usepackage{color}
\usepackage{soul}

\newcommand{\apgt}{{\raise-.5ex\hbox{$\buildrel>\over\sim$}}}
\newcommand{\aplt}{{\raise-.5ex\hbox{$\buildrel<\over\sim$}}} 
\newcommand{\Mpc}{\rm\; Mpc}

\newcommand{\km}{\rm\; km}

\newcommand{\s}{\rm\; s}

\newcommand{\keV}{\rm\; keV}

\newcommand{\ergps}{\hbox{$\erg\s^{-1}\,$}}

\newcommand{\kmps}{\hbox{$\km\s^{-1}\,$}}

\newcommand{\kmpspMpc}{\hbox{$\kmps\Mpc^{-1}\,$}}

\newcommand{\Omm}{\hbox{$\rm\thinspace \Omega_{m}$}}
\newcommand{\OmL}{\hbox{$\rm\thinspace \Omega_{\Lambda}$}}

\def\aap{A\&A}                   

\def\apjs{ApJS}                  
\def\apjl{ApJ}                   

\def\ltsim{\mathrel{\hbox{\rlap{\hbox{\lower4pt\hbox{$\sim$}}}\hbox{$<$}}}}
\def\gtsim{\mathrel{\hbox{\rlap{\hbox{\lower4pt\hbox{$\sim$}}}\hbox{$>$}}}}

\def\ergps{ erg s$^{-1}$}

\def\aap{A\&A}

\def\apjs{ApJS}

\def\apjl{ApJL}

\title[]{$Chandra$ X-ray observations of the hyper-luminous infrared galaxy IRAS F15307+3252}

\author[J. Hlavacek-Larrondo, P. Gandhi, M.T. Hogan et al.]{J. Hlavacek-Larrondo$^{1}$\thanks{E-mail: juliehl@astro.umontreal.ca}, P. Gandhi$^{2}$, M.T. Hogan$^{3,4}$, M.-L. Gendron-Marsolais$^{1}$, \newauthor{A.C. Edge$^5$, A.C. Fabian$^6$, H.R. Russell$^6$, K. Iwasawa$^{7,8}$, M. Mezcua$^{1}$}\\
$^{1}$D\'{e}partement de Physique, Universit\'{e} de Montr\'{e}al, Montr\'{e}al, QC H3C 3J7, Canada\\
$^{2}$School of Physics $\&$ Astronomy, University of Southampton, Highfield, Southampton, SO17 1BJ, UK\\
$^{3}$Department of Physics and Astronomy, University of Waterloo, 200 University Avenue West, Waterloo, ON N2L 3G1, Canada\\
$^{4}$Perimeter Institute for Theoretical Physics, Waterloo, ON N2L 2Y5, Canada\\
$^{5}$Institute of Computational Cosmology, Department of Physics, Durham University, Durham, DH1 3LE, UK\\
$^{6}$Institute of Astronomy, University of Cambridge, Madingley Road, Cambridge CB3 0HA\\
$^{7}$Institut de Ci\`encies del Cosmos (ICCUB), Universitat de Barcelona (IEEC-UB), Mart\'i i Franqu\`es, 1, 08028 Barcelona, Spain\\
$^{8}$ICREA, Pg. Llu\'is Companys 23, 08010 Barcelona, Spain\\}

\begin{document}
\date{}

\pagerange{\pageref{firstpage}--\pageref{lastpage}} \pubyear{2016}

\maketitle

\begin{abstract}
\noindent Hyper-luminous infrared galaxies (HyLIRGs) lie at the extreme luminosity end of the IR galaxy population with $L_{\rm IR}>10^{13}$L$_\odot$. They are thought to be closer counterparts of the more distant sub-mm galaxies, and should therefore be optimal targets to study the most massive systems in formation. We present deep $Chandra$ observations of IRAS~F15307+3252 (100ks), a classical HyLIRG located at $z=$0.93 and hosting a radio-loud AGN ($L_{\rm 1.4~GHz}\sim3.5\times10^{25}$~W/Hz). The $Chandra$ images reveal the presence of extended ($r=160$~kpc), asymmetric X-ray emission in the soft 0.3-2.0~keV band that has no radio counterpart. We therefore argue that the emission is of thermal origin originating from a hot intragroup or intracluster medium virializing in the potential. We find that the temperature ($\sim2$~keV) and bolometric X-ray luminosity ($\sim3\times10^{43}$~erg s$^{-1}$) of the gas follow the expected $L_{\rm X-ray}-T$ correlation for groups and clusters, and that the gas has a remarkably short cooling time of $1.2$~Gyrs. In addition, VLA radio observations reveal that the galaxy hosts an unresolved compact steep-spectrum (CSS) source, most likely indicating the presence of a young radio source similar to 3C186. We also confirm that the nucleus is dominated by a redshifted 6.4~keV Fe K$\alpha$ line, strongly suggesting that the AGN is Compton-thick. Finally, Hubble images reveal an over-density of galaxies and sub-structure in the galaxy that correlates with soft X-ray emission. This could be a snapshot view of on-going groupings expected in a growing cluster environment. IRAS~F15307+3252 might therefore be a rare example of a group in the process of transforming into a cluster.
\end{abstract}




\section{Introduction}
Hyper-luminous infrared galaxies (HyLIRGS) are the most luminous and highly star-forming objects in the Universe with infrared (IR) luminosities in excess of $10^{13}$L$_\odot$ \citep[e.g.][]{Row2000316}. The redshifts of these sources coincide with the epoch when star-formation was peaking and the largest cosmic structures were maturing \citep[$z\sim1$;][]{Cha2001556}. HyLIRGs are thought to be closer counterparts of the more distant sub-mm galaxies \citep[e.g.][]{Cas2014541}, so they ought to be optimal targets to study the most massive, luminous systems in formation. Their enormous radiative power outputs require them to be powered in large by radiatively-efficient accretion onto the central black hole (Lutz et al. 1998, Ruiz et al. 2010, Nandra \& Iwasawa 2007). However, we still do not understand this epoch of black hole growth and how the growth of black holes is connected to their host galaxies or the larger-scale environment. We can start tackling these questions at X-ray wavelengths since X-rays provide a means to trace emission from Active Galactic Nuclei (AGN), as well as from hot diffuse gas surrounding galaxies. However, many HyLIRGs turn out to be faint or undetected in X-rays (e.g. Wilman et al. 2003), probably as a result of extreme obscuration, yet there are a few exceptions of X-ray emitting HyLIRGs, one of which is the target of this paper.

IRAS F15307+3252 is a classical HyLIRG located at a redshift of $z$=0.93 \citep[e.g.][]{Iwa2005362,Dea2001326,Fab1996283}. It is also a luminous ($M_I=-26.4$), giant elliptical galaxy with a bolometric luminosity of 3$\times10^{13}$L$_\odot$ that shows no evidence of lensing from Hubble Space Telescope ($HST$) images \citep[][]{Far2002329}. There is also no detection of molecular gas associated with the galaxy based on IRAM 30-m observations \citep[M$_{\rm H_2}<7.1\times10^{9}M_\odot$;][]{Com2013550}. 

However, in several short ($\ltsim$10 ks) exposures with $XMM-Newton$, Iwasawa et al. (2005) uncovered two distinct components in the X-ray spectrum. First, a significant excess of counts ($5\sigma$) at the redshifted energy of the 6.4 keV Fe K$\alpha$ line was detected at hard energies, superposed on a faint and very hard continuum over the $2-10$ keV range. An estimate of the intrinsic AGN continuum needed to produce the observed Fe line flux showed that the source must have a quasar luminosity of $L_{\rm{2-10~keV}}\sim 10^{45}$ erg s$^{-1}$. Assuming a typical quasar spectral energy distribution, this means that the AGN can easily power more than 30 \% of the bolometric luminosity which is estimated to be on the order of $10^{47}$ erg s$^{-1}$. These results are consistent with the source being a Compton-thick AGN dominated by reflection (e.g., Leahy \& Creighton 1993).

\citet{Iwa2005362} also found evidence of an upturn in the spectrum below 2 keV, which appeared to be extended on scales larger than the resolution limit of $XMM-Newton$ with a full-width at half maximum (FWHM) of 21$\pm$5 arcsec or $60-100$ kpc in radius. Fitting a thermal plasma model, the authors found that the temperature of the soft X-ray emission is $2.1_{-0.4}^{+0.6}$ keV and that the X-ray bolometric luminosity is $\sim10^{44}$ erg s$^{-1}$. These properties are consistent with the standard $L_{\rm{X-ray}}-T$ relation of clusters \citep[e.g.][]{Sta2006648}, implying that IRAS F15307+3252 is surrounded by a hot medium similar to that expected for a Virgo-like cluster. Thus, IRAS F15307+3252 probably represents an early phase in the growth of a giant elliptical galaxy such as M~87 as well as its host cluster. 

Here we report the results from a 100ks $Chandra$ observation of IRAS F15307+3252, which confirms the presence of extended X-ray emission surrounding the galaxy, although the resolution of $Chandra$ reveals that it is highly asymmetric. This example provides a unique opportunity to study the heating and cooling balance between a black hole and its surrounding environment at an epoch still unexplored. In Section 2, we describe the observations, and then in Section 3 we present our main results. In Section 4, we discuss the implications of these results. Finally, in Section 5 we state our conclusions. Throughout this paper, we adopt $H_\mathrm{0}=70\kmpspMpc$ with $\Omm=0.3$, $\OmL=0.7$, resulting in a scaling of 7.862 kpc per arcsec. All errors are $2\sigma$ unless otherwise noted.

\section{Observations and data reduction}

\subsection{$Chandra$ X-ray observations}

\begin{figure}
\centering
\begin{minipage}[c]{0.99\linewidth}
\centering \includegraphics[width=\linewidth]{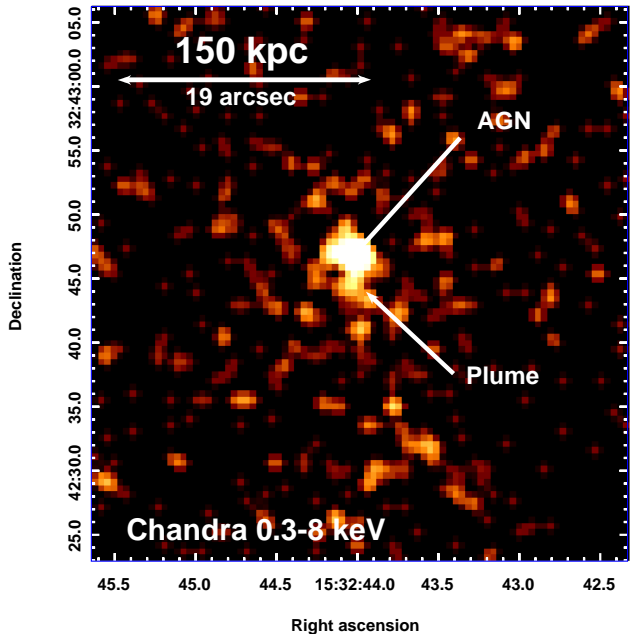}
\end{minipage}
\caption[]{Exposure-corrected $0.3-8$ keV $Chandra$ image of IRASF15307+3252 totalling 98.3 ks, spatially smoothed with a 2 arcsec gaussian function and covering an area of 330 kpc by 330 kpc. We highlight the AGN in the galaxy and a small, plume-like feature to the south.}
\label{fig1}
\end{figure}

IRAS F15307+3252 was observed in 2012 with the $Chandra$ CCD Imaging Spectrometer (ACIS) in VFAINT mode and centred on the ACIS-S3 back-illuminated chip (100 ks; ObsID 13907) with the ACIS-S1, ACIS-S2, ACIS-S4 and ACIS-I3 also switched on. The observations were processed, cleaned and calibrated using {\sc ciao} ({\sc ciaov4.5}, {\sc caldb 4.5.6}), and starting from the level 1 event file. We applied both charge transfer inefficiency and time-dependent gain corrections, as well as removed flares using the {\sc {lc$\_$clean}} script with a $3\sigma$ threshold. The net exposure time is shown in Table \ref{tab1}. We then exposure-corrected the image, using an exposure map generated with a monoenergetic distribution of source photons at 1.5 keV, which is the energy expected for the thermal emission as seen by \citet{Iwa2005362}. The resulting exposure-corrected $0.3-8.0\keV$ image is shown in Fig. \ref{fig1}. We highlight the central AGN of the galaxy, as well as a small, extended plume-like feature to the south.  

\begin{table}
\caption[]{$Chandra$ X-ray observations.}
\begin{tabular}{lccc}
\hline
Observation number & Date & Detector & Exposure (ks) \\
\hline
13907 & 07/11/2012& ACIS-S & 98.3 \\
\hline
\end{tabular}
\label{tab1}
\end{table}
 
The X-ray data were spectroscopically analysed with {\sc xspec} (v12.9.0o). We use the abundance ratios of \citet{And198953} throughout this study, as well as C-statistics. The background was in most cases selected as a region located on the same chip, devoid of point sources, and far from any emission associated with IRAS F15307+3252. IRAS F15307+3252 occupies only a small fraction of the chips, and a background can easily be extracted from the surrounding region. Note that our choice of background does not significantly affect our results. If we use blank-sky observations or select a similar region located on the second back-illuminated chip (ACIS-S1), we find results consistent with each other. We use the \citet{Kal2005440} value for the Galactic absorption throughout this study (N$_{\rm H}=2.19\times10^{20}$cm$^{-2}$). 

\begin{table*}
\caption[]{VLA radio observations.}
\resizebox{18cm}{!} {
\begin{tabular}{lccccccc}
\hline
Project & Frequency & Bandwidth & Configuration & beam & rms & Peak flux & Integrated Flux \\
\hline
AB0654	& X-band (8.4 GHz)	& 	50 MHz	&  D	& $9.8\times8.1$ arcsec$^2$ &  0.05 mJy/beam & 0.81$\pm$0.05 mJy/beam & 1.0$\pm$0.1 mJy\\
AG0514	& X-band (8.4 GHz)	& 	50 MHz	&  B	& $0.8\times0.7$ arcsec$^2$ &  0.01 mJy/beam & 0.92$\pm$0.01 mJy/beam & 1.02$\pm$0.05 mJy\\
\hline
NVSS    & L-band (1.4 GHz) 	& 	50 MHz	&  D$\&$DnC	& $45\times45$ arcsec$^2$ &  0.45 mJy/beam &  7.4 mJy/beam& 8.1$\pm$0.5 mJy\\
FIRST    & L-band (1.4 GHz) 	& 	50 MHz	&  B	& $5.9\times5.0$ arcsec$^2$ &  0.1 mJy/beam & 5.7 mJy/beam & 5.9 mJy\\
AB0806	& L-band (1.4 GHz)	& 	50 MHz	&  A	& $1.5\times1.4$ arcsec$^2$ &  0.1 mJy/beam & 5.7$\pm$0.1 mJy/beam & 6.0$\pm$0.4 mJy\\
\hline
WENSS	& 326 MHz &   5.0 MHz	& ...&  $54\times54$ arcsec$^2$ &  3.4 mJy/beam & 18.0$\pm$3.4 mJy/beam & 23.0$\pm$3.4 mJy\\
\hline
TGSS	& 150 MHz & 	 16 MHz	& ...& $25\times25$ arcsec$^2$ &  4 mJy/beam & 11$\pm$4 mJy/beam & 11$\pm$4 mJy\\
\hline
\end{tabular}}
\label{tab2}
\end{table*}

\subsection{VLA observations}

IRAS F15307+3252 was observed on three occasions with the VLA from 1992 to 1996: projects AB0654 (6 minutes on source), AG0514 (4 hours on source) and AB0806 (3 minutes on source). The data were reduced using the standard Astronomical Image Processing System\citep[\textsc{aips};][]{Gre2003285}. All resulting values are quoted in Table \ref{tab2}. A point source was detected at the location of the galaxy for each dataset. No extended emission was found. Note that the integrated values for the three datasets include a 5$\%$ absolute flux error. 

The source is not detected in the VLA Low Frequency Sky Survey \citep[VLSS, rms$\sim0.1$ Jy/beam;][]{Coh2007134}. However, the reprocessed 150 MHz images obtained with the alternative TIFR GMRT Sky Survey \citep[TGSS\footnote{http://tgssadr.strw.leidenuniv.nl/doku.php};][]{Int2016} reveal the presence of a faint point source at the location of the galaxy, although it falls below the detection limit of the catalogue. The 326 MHz images taken with the Westerbork Northern Sky Survey \citep[WENSS;][]{Ren1997124} also reveal a faint point source at the location of IRAS F15307+3252. We include the values from both these surveys in Table \ref{tab2}, as well as the 1.4 GHz VLA Faint Images of the Radio Sky at Twenty-cm survey  \citep[FIRST;][]{Bec199461} and the 1.4 GHz NRAO VLA Sky Survey catalogue \citep[NVSS;][]{Con1998115}.

\begin{figure*}
\centering
\begin{minipage}[c]{0.6\linewidth}
\centering \includegraphics[width=\linewidth]{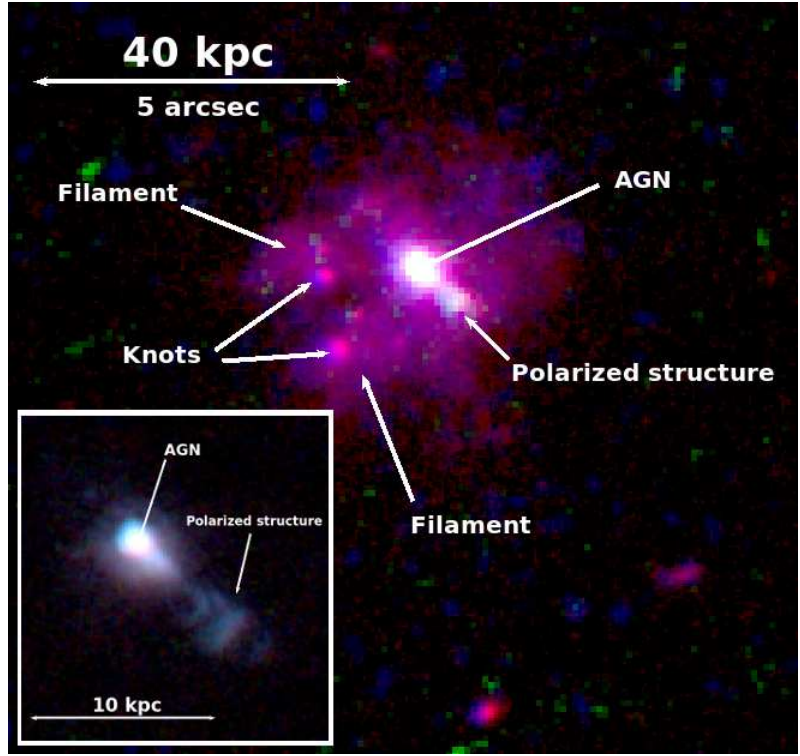}
\end{minipage}
\caption[]{Combined $HST$ ACS/WFC F814W (red), WFPC2 F791W (blue) and WFPC2 F450W (green) images of IRAS F15307+3252. The inset to the bottom-left shows the $HST$ polarisation images of the central regions of the galaxy. These were obtained with ACS/HRC using the F606W filter. We highlight the extended polarized structure to the south-west.}
\label{fig2}
\end{figure*}

\subsection{Hubble Space Telescope observations}
IRAS F15307+3252 was observed with $HST$ several times from 1997 to 2005. Fig. \ref{fig2} shows a combined image taken with the Wide-Field Planetary Camera 2 (WFPC2) in two filters (F450W and F791W) and the Advanced Camera for Surveys (ACS) Wide Field Channel (WFC) in the F814W filter. The F791W filter contains the [$O_{\rm{II}}$]$\lambda$3732 line, which traces hot gas kinematics such as ionization cones. In Fig. \ref{fig2}, we also show the combined ACS High Resolution Channel (HRC) images obtained with the F606W filter, applying a POL0V, POL60V and POL120V polarisation filter. These highlight the polarized structures in the galaxy, especially the cone-like feature to the south-west. This feature also appears strongly in the F791W image. 

\section{Results}

\subsection{Extended X-ray emission}

\begin{figure*}
\centering
\begin{minipage}[c]{0.99\linewidth}
\centering \includegraphics[width=\linewidth]{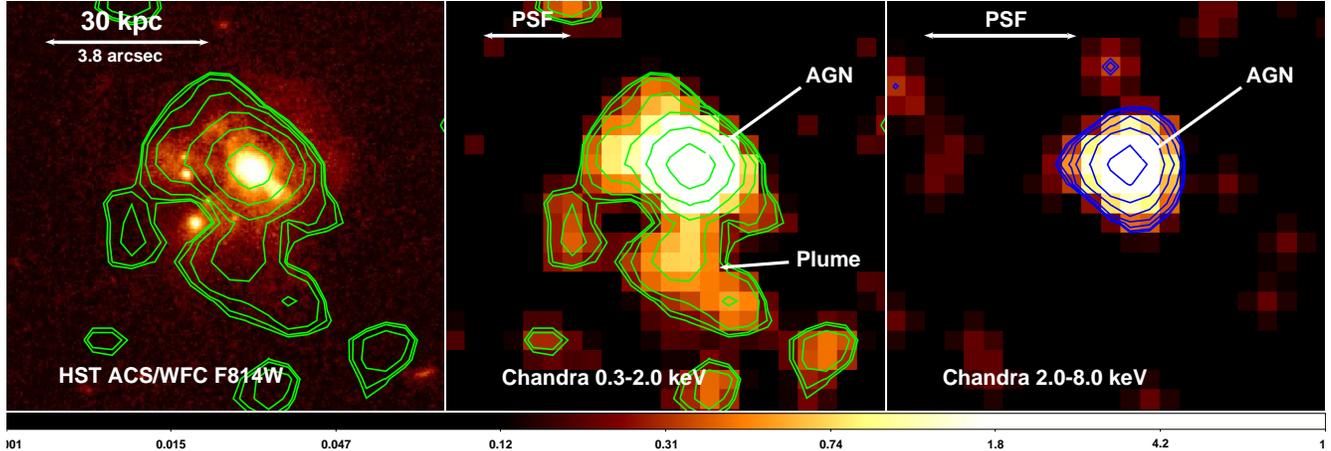}
\end{minipage}
\caption[]{Small-scale image of IRAS F15307+3252. Left: $HST$ ACS/WFC image of the central regions of IRAS F15307+3252 obtained with the F814W filter. Middle: $Chandra$ image in the soft X-ray band ($0.3-2.0$ keV), smoothed with a gaussian function of 2 arcsec. The green contours highlight the detected emission in the $0.3-2.0$ keV band, starting at a 2$\sigma$ level. These contours are also overplotted in the left panel. They highlight the same small, plume-like extended emission to the south as in Fig. 1. Right: $Chandra$ image in the hard X-ray band ($2.0-8.0$ keV), smoothed with a gaussian function of 2 arcsec and showing the AGN point source. The blue contours show the hard X-ray band emission ($2.0-8.0$ keV), also starting at a 2$\sigma$ level. The color bar indicates the number of counts per pixel for the $0.3-2.0$ keV image.}
\label{fig3}
\end{figure*}

In Fig. \ref{fig3}, we show the $HST$ WFC image of the central regions of IRAS F15307+3252 obtained with the F814W filter in the left panel. The middle and right panels show the $Chandra$ $0.3-2.0$ keV and $2.0-8.0$ keV images, respectively. We overplot in green, contours for any soft $0.3-2.0$ keV emission detected, starting at a 2$\sigma$ level. The blue contours show the equivalent, but for the hard $2.0-8.0$ keV band. Also shown is the point spread function (PSF) of the $Chandra$ telescope, for each of the energy bands. Note that the PSF FWHM of the $XMM-Newton$ detectors used in \citet{Iwa2005362} is 6 arcsec. The green contours clearly reveal an extended plume-like structure in the soft band that is not seen in the hard band. This plume was also seen in the $0.3-8.0$ keV exposure-corrected image in Fig. \ref{fig1}. It appears to first extend in the southern direction, and then curves towards the west with a maximum extension of $r=35$ kpc. Fig. \ref{fig3} also shows that there is extended soft X-ray emission to the east of the galaxy that correlates with the eastern filament seen in the optical band and highlighted in Fig. \ref{fig2}. Archival $Spitzer$ images of the source, although of poorer resolution, show no clear association with the extended X-ray emission. 

\begin{figure*}
\centering
\begin{minipage}[c]{0.32\linewidth}
\centering \includegraphics[width=\linewidth]{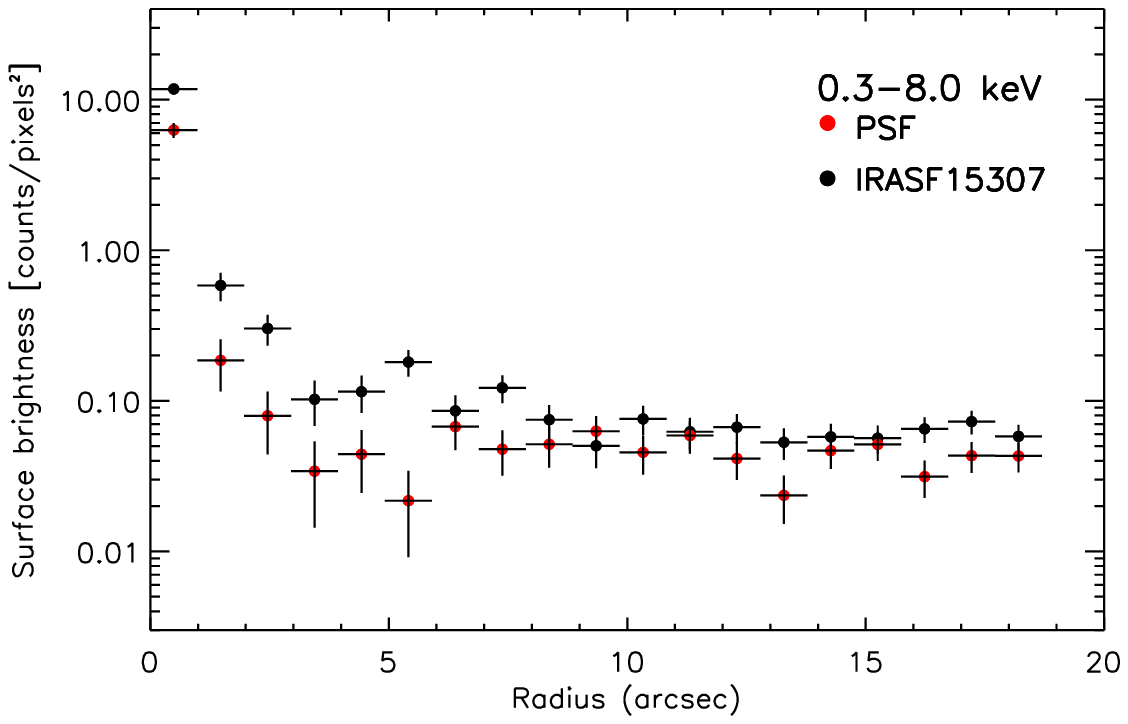}
\end{minipage}
\begin{minipage}[c]{0.32\linewidth}
\centering \includegraphics[width=\linewidth]{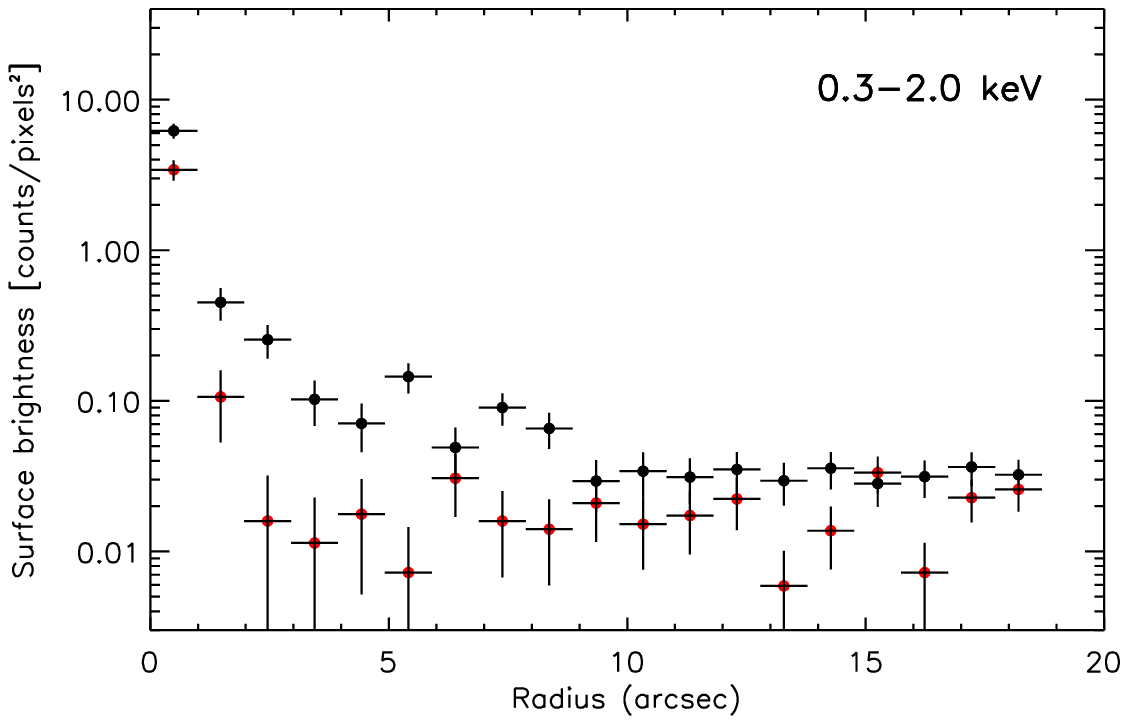}
\end{minipage}
\begin{minipage}[c]{0.32\linewidth}
\centering \includegraphics[width=\linewidth]{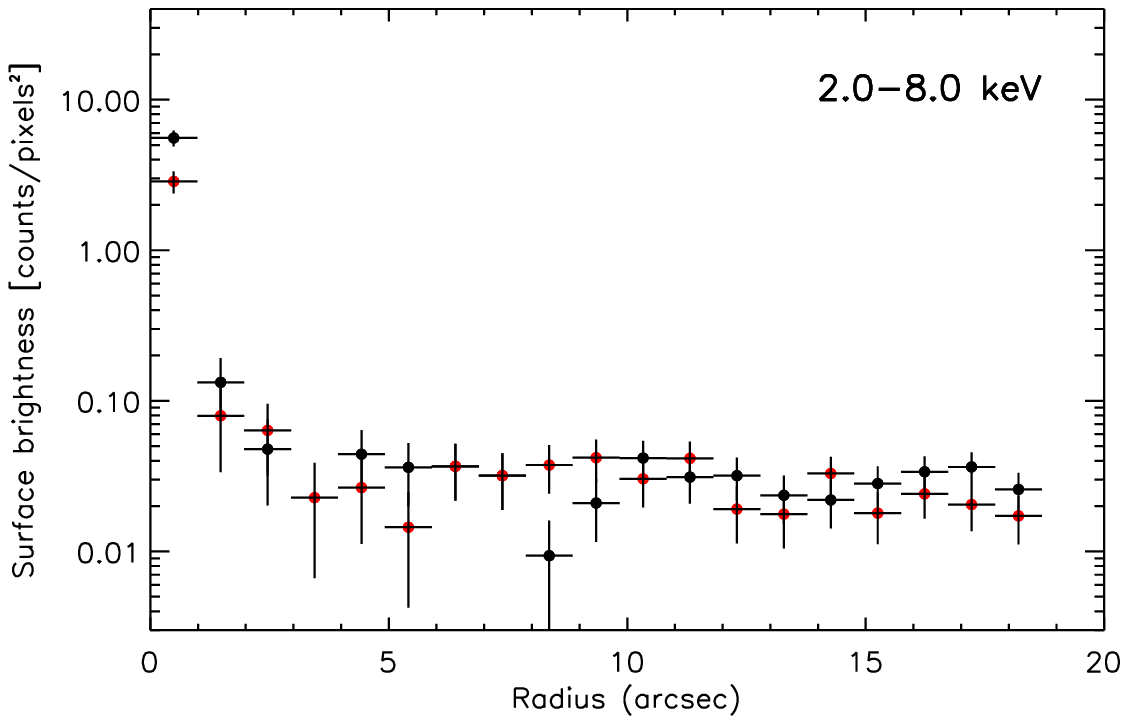}
\end{minipage}
\caption[]{Left: Surface brightness profile in the $0.3-8.0$ keV energy range of the $Chandra$ PSF (red) and the observed emission (black). Middle: Same but for the $0.3-2.0$ keV energy range. Right: Same but for the $2.0-8.0$ keV energy range. These plots show that IRAS F15307+3252 is indeed extended at X-ray wavelengths, but only in the soft $0.3-2.0$ keV band.}
\label{fig3b}
\end{figure*}

To further demonstrate the existence of extended emission, we simulated observations of a point-like AGN using the ChaRT $Chandra$ ray-tracing program (Carter et al. 2003), and the MARX software (v5) to project the rays on to the ACIS-S detector. We provided as  inputs  the  position  of  the  AGN (RA=15:32:44.03, DEC=+32:42:46.75),  the exposure time (98.3 ks) and the quasar spectrum determined using the surrounding annulus as a background (model II, see Section 3.2). The resulting PSF profiles (red), compared to those observed (black), are shown in Fig. \ref{fig3b} for three energy bands. This figure clearly shows that the source is extended in the soft $0.3-2.0$ keV band and that the extended emission dominates over the quasar emission beyond 1 arcsec.

The net X-ray counts of the extended X-ray emission in the $0.3-2.0$ keV band, calculated by extracting the counts within an outer annulus of 4.4 arcsec (35 kpc), where the emission is detected at at least 2$\sigma$ significance (Fig. \ref{fig3}), and an inner radius of 1 arcsec (equivalent to the PSF), is 58. The southern part of this extended X-ray emission (i.e. excluding the eastern extension) accounts for half of the counts. If we include the nuclear emission, we find that within a radius of 4.4 arcsec there are 123 X-ray counts. The nucleus therefore accounts for roughly half of the counts within $r<35$ kpc. 

\begin{table}
\caption[]{Spectral modelling for the extended X-ray emission in IRAS F15307+3252. (1) Model name; (2) power law index; (3) flux density at 1 keV if the model includes a power law; (4) temperature of the \textsc{apec} model assuming an abundance of 0.3 Z$_\odot$; (5) unabsorbed $0.3-8.0$ keV X-ray luminosity (observed frame). Galactic absorption was frozen at the value of Kalberla et al. (2005). The first section of the table used a region located at 7.8 kpc $<r< 35$ kpc, while the second includes the $r\sim160$ kpc extended emission seen in Fig. \ref{fig4}. }
\begin{tabular}{lcccc}
\hline
 Model name  & $\Gamma$ & F$_\nu$ & kT &  L$_{\rm{0.3-8.0 keV}}$ \\
 & & [nJy] & [keV] & [10$^{43}$ erg s$^{-1}$] \\
\hline
phabs [pow] & 2.7$\pm0.6$ & 61$^{+20}_{-16}$  & ... & 1.9$^{+0.7}_{-0.5}$ \\
phabs [apec] & ... & & 1.7$^{+1.2}_{-0.4}$ &  1.5$^{+0.5}_{-0.4}$ \\
\hline
phabs [pow] & 2.6$^{+0.6}_{-0.5}$ & 93$^{+27}_{-24}$  & ... & 2.9$^{+0.9}_{-0.8}$ \\
phabs [apec] & ... & & 2.2$^{+1.7}_{-0.6}$ &  2.3$^{+0.7}_{-0.6}$ \\
\hline
\end{tabular}
\label{tab4}
\end{table}

\begin{figure*}
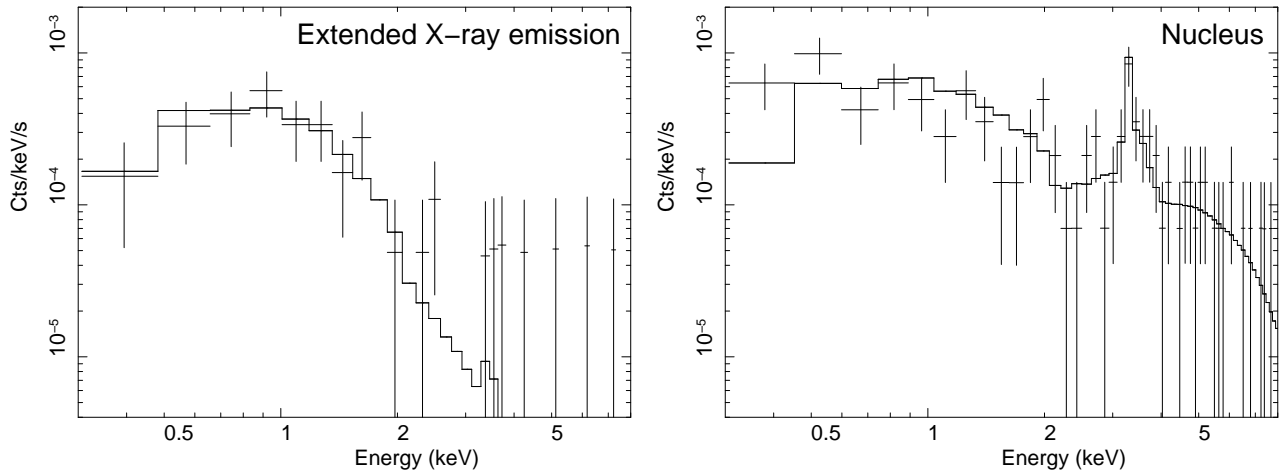

\centering
\begin{minipage}[c]{0.48\linewidth}
\centering \includegraphics[angle=270,width=\linewidth]{./Thermal.ps}
\end{minipage}
\begin{minipage}[c]{0.48\linewidth}
\centering \includegraphics[angle=270,width=\linewidth]{./Nucleus.ps}
\end{minipage}
\caption[]{We show the spectrum of the extended X-ray emission surrounding the nucleus at $7.8<r<35$ kpc (left panel) and the spectrum of the nucleus at $r<7.8$ kpc (right panel). Both include models fit to the data (solid line). In the first case, we fit a thermal plasma model (\textsc{apec}) and in the second, we fit a thermal model including a torus component (\textsc{apec + torus}). The latter shows the best fitting value of the absorption (N$_{\rm H}\sim20\times10^{24}$ cm$^{-2}$), although the fit is consistent with a lower limit only. See Section 3 for details. }
\label{fig5}
\end{figure*}

Using the same region excluding the nucleus, we extracted an X-ray spectrum and fit different models to the emission in the $0.3-8.0$ keV energy range. The results are summarised in Table \ref{tab4}. First, we fit an absorbed power law model with the absorption fixed to the Galactic value of \citet{Kal2005440}, that is N$_{\rm H}=2.19\times10^{20}$cm$^{-2}$. The background was selected as a region located far from the galaxy, devoid of point sources. We let the power law index free to vary, as well as the normalisation. We find that the fit is well constrained, with a power law index of $\Gamma=2.7\pm0.6$ and a $0.3-8.0$ keV flux density of 61$^{+20}_{-16}$ nJy at 1 keV. The emission can also be fitted with a thermal plasma model (\textsc{apec}), assuming a fixed metallicity of 0.3 solar. This is a typical value for clusters, even at high-$z$ \citep[e.g.][]{McD2016a}. Changing the metallicity to 0.2 solar does not affect our results. This yields a temperature of $kT=1.7^{+1.2}_{-0.4}$ keV, and an X-ray luminosity in the $0.3-8.0$ keV range of $1.5^{+0.5}_{-0.4}\times10^{43}$ \ergps. The spectrum of the emission within $r=35$ kpc, excluding the nucleus ($r<1$ arcsec), is shown in the left panel of Fig. \ref{fig5}.

\begin{figure}
\centering
\begin{minipage}[c]{0.99\linewidth}
\centering \includegraphics[width=\linewidth]{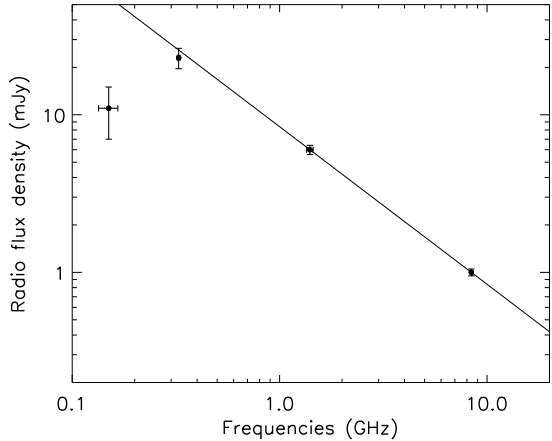}
\end{minipage}
\caption[]{Radio spectral energy distribution of the AGN in IRAS F15307+3252. The solid line shows the expected slope for emission following a power law with $\alpha=1$, where $S_\nu\propto\nu^{-\alpha}$. The compactness of the source, and the turnover at low frequencies ($\sim200$ MHz), suggests that the radio source in IRAS F15307+3252 is a CSS source. }
\label{figsed}
\end{figure}

The radio images obtained in Section 2 do not reveal any radio counterpart associated with the extended X-ray emission, although we find an unresolved radio point source that coincides with the location of the AGN. The spectral index ($\alpha$) of this point source, defined such that the flux density scales as $S_\nu\propto\nu^{-\alpha}$, is $\alpha\sim1$ between 1.4 GHz and 8.4 GHz. We show in Fig. \ref{figsed} the radio spectral energy distribution (SED) of the source. This plot reveals that the emission at low frequencies falls below the expected emission for $\alpha\sim1$ (solid line). The turnover at $\sim200$ MHz, and the compact nature of the source, therefore strongly suggests that the galaxy harbours a compact steep-spectrum (CSS) source at its core.

\begin{figure*}
\centering
\begin{minipage}[c]{0.99\linewidth}
\centering \includegraphics[width=\linewidth]{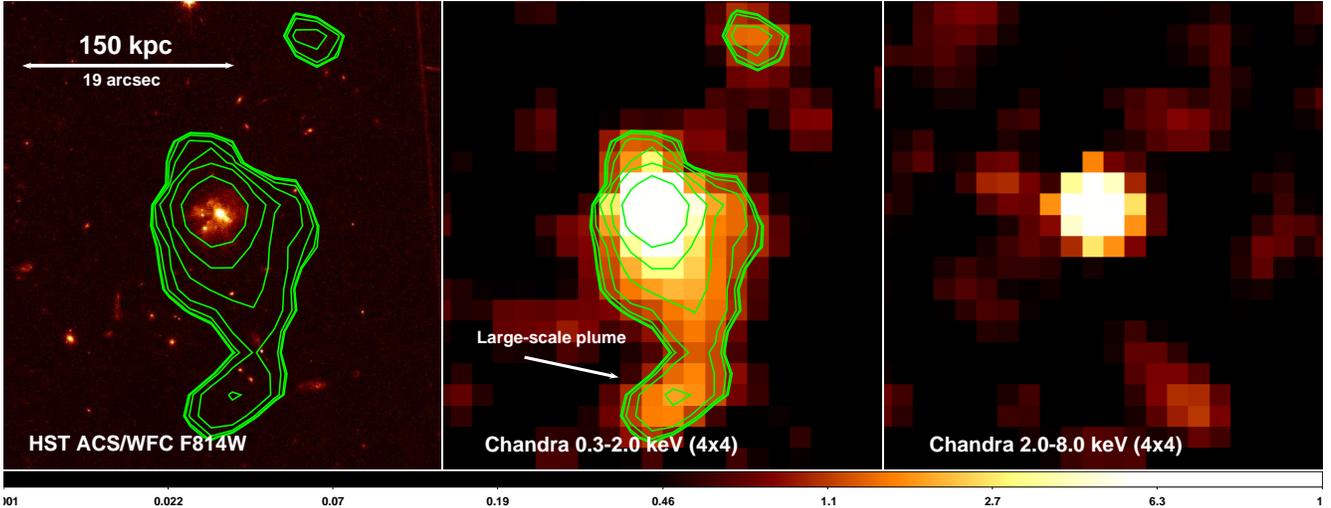}
\end{minipage}
\caption[]{Large-scale image of IRAS F15307+3252, covering the same area as in Fig. \ref{fig1} (330 kpc by 330 kpc). Left: $HST$ ACS/WFC image of IRAS F15307+3252 obtained with the F814W filter. Middle: $Chandra$ image in the soft X-ray band ($0.3-2.0$ keV) binned by 4$\times$4, smoothed with a gaussian function of 2 arcsec. This image reveals large-scale, extended X-ray emission to the south. Green contours highlight this large-scale X-ray plume with contours starting at 2$\sigma$, and are overplotted in the left panel as well. Right: $Chandra$ image in the hard X-ray band ($2.0-8.0$ keV), showing the AGN point source and no extended X-ray emission. The color bar indicates the number of counts per pixel for the $0.3-2.0$ keV image.}
\label{fig4}
\end{figure*}

In Fig. \ref{fig4}, we focus on the large-scale properties of IRAS F15307+3252 and its surroundings. We show the large-scale $HST$ ACS/WFC F814W image in the left panel. We note that there is a chain of galaxies located $\sim13$ arcsec to the south of IRAS F15307+3252 that may be associated with IRAS F15307+3252 \citep[see e.g.][]{Iwa2005362,Fab1996283}, but these could also be galaxies seen in projection. The middle panel shows the $Chandra$ image in the $0.3-2.0$ keV band on the same scale, binned by a factor of 4, such that each pixel corresponds to 16 ($4\times4$) pixels in the original image. The right panel shows the $2.0-8.0$ keV band, also on the same scale and binned by a factor of 4. Similarly to Fig. \ref{fig3}, green contours trace the soft X-ray emission and are overplotted on the $HST$ image, with contours starting at 2$\sigma$. This image reveals that the small-scale plume seen in Fig. \ref{fig3}, may extend out to a radius of 160 kpc, in agreement with \citet{Iwa2005362}. It is highly asymmetric and has no counterpart to the north.

This large-scale X-ray plume contains 96 counts, compared to 143 counts if we include the nucleus in the soft X-ray band. Similarly to the small-scale plume, we fit different models to the emission in the $0.3-8.0$ keV energy range (see Table \ref{tab4}), excluding the central nucleus in a 4 arcsec circular region. An absorbed power law model with the absorption fixed to the Galactic value produces a good fit, with a power law index of $\Gamma=2.6\pm^{+0.6}_{-0.5}$ and a $0.3-8.0$ keV flux density of 93$^{+27}_{-24}$ nJy at 1 keV. These values are similar to those found for the small-scale plume in Fig. \ref{fig3}, although somewhat larger. A thermal plasma model (\textsc{apec}) also produces a good fit, assuming a fixed metallicity of 0.3 solar. We obtain a temperature of $kT=2.2^{+1.7}_{-0.6}$ keV, and an X-ray luminosity in the $0.3-8.0$ keV range of $2.3^{+0.7}_{-0.6}\times10^{43}$ \ergps.

Overall, Figs. \ref{fig3} and \ref{fig4} both reveal the presence of extended X-ray emission surrounding IRAS F15307+3252, that may extend up to $r=160$ kpc in the southern direction. We discuss these results in Section 4.

\subsection{X-ray emission from the AGN}

We extract a $0.3-8.0$ keV spectrum within a circular region of radius 1 arcsec, centred on the AGN as seen from the hard $2-8$ keV image (Fig. \ref{fig3}). We then take a surrounding region located far from the galaxy and devoid of any X-ray point sources as a background. All of our fits include Galactic absorption, which we keep frozen at the Kalberla et al. (2005) value as in the previous section.  

In Table \ref{tab3} we present the results from the various spectral models that were fit to the data for the AGN. The first model fits the spectrum with a simple absorbed (Galactic) power law and Gaussian line (model I). We include the Gaussian line since it is evident in the spectrum and was already known from past studies \citep[e.g.][]{Iwa2005362}. Fitting the data, we find that the fit tends to produce an unreasonable value for the central line energy and dispersion of the Gaussian line. Instead, we keep the line energy frozen to 3.3 keV which is simply the value for the redshifted 6.4 Fe K$\alpha$ line. We also show the results while keeping the power law index frozen to 1.8 (model II). The results, including the unabsorbed $2-10$ keV (rest-frame) luminosities are shown in Table \ref{tab3}. We also tested an internally absorbed \textsc{phabs[zphabs(power + gaussian)]} model, but the fit favoured no internal absorption ($N_{\rm{H}}=0$). We attempted to fit the data using a Compton reflection component from neutral matter \citep[\textsc{pexrav};][]{Mag1995273} but the fit remained largely unconstrained, most likely due to the small number of counts. 

\begin{table*}
\caption[]{Spectral modelling for the AGN in IRAS F15307+3252. (1) Model number; (2) model name; (3) absorbing column density at the redshift of the source; (4) power law index; (5) Gaussian central energy; (6) Gaussian dispersion; (7) line luminosity in the $2-10$ keV range; (8) rest-frame, unabsorbed non-thermal $2-10$ keV X-ray luminosity (including the line emission). Galactic absorption was frozen at the value of Kalberla et al. (2005). For model III and IV, we keep the temperature and abundance frozen at $kT=1.7$ and Z=0.3Z$_\odot$, respectively. The first section of the table uses a background located far from the galaxy, while the second uses a local background immediately surrounding the AGN. See Section 3 for details.}
\begin{tabular}{llcccccc}
\hline
Model number & Model name & N$_{\rm{H}}$ & $\Gamma$ & E$_{\rm{gaussian}}$ & $\sigma_{\rm{gaussian}}$ & L$_{\rm{line, 2-10 keV}}$ & L$_{\rm{2-10 keV}}$ \\
 & & [10$^{22}$ cm$^{-2}$] &  & [keV] & [keV] &  [10$^{43}$ erg s$^{-1}$] & [10$^{43}$ erg s$^{-1}$] \\
\hline
I	& phabs [pow+ga] & ... & 1.5$^{+0.4}_{-0.3}$ & 3.3 (fr)& 0.23$^{+0.25}_{-0.21}$ & 1.4$^{+1.0}_{-0.6}$& 4.4$^{+2.0}_{-1.4}$ \\
II	& phabs [pow+ga] & ... & 1.8 (fr) & 3.3 (fr)& 0.30$^{+0.33}_{-0.16}$ & 1.7$^{+0.9}_{-0.8}$& 4.2$^{+1.4}_{-1.2}$ \\
III & phabs [apec + zphabs(pow+ga)] & 30$^{+33}_{-14}$ & 1.8 (fr) & 3.3 (fr) & 0.3 (fr)  & 1.9$^{+2.8}_{-1.4}$ & 11.7$^{+11.3}_{-5.5}$ \\ 
IV & phabs [apec + torus] & $>250$ & 1.7$^{+0.6}_{-0.7}$  & ... & ...  & ... & $\sim230$ \\
\hline
I	& phabs [pow+ga] & ... & 3.2$^{+0.7}_{-0.6}$ & 3.3 (fr) & 0.35$^{+0.60}_{-0.20}$ & 2.3$^{+1.6}_{-1.0}$& 2.9$^{+2.1}_{-1.3}$ \\
II	& phabs [pow+ga] & ... & 1.8 (fr) & 3.3 (fr) & 0.29$^{+0.23}_{-0.18}$ & 1.7$^{+1.0}_{-0.9}$&  3.6$^{+1.5}_{-1.3}$ \\
IV & phabs [apec + torus] & $>350$ & 1.8 (fr)  & ... & ...  & ... &  $\sim190$ \\
\hline
\end{tabular}
\label{tab3}
\end{table*}

Next, we fit the spectrum of the nucleus with an absorbed thermal plasma model (\textsc{apec}), including a power law and Gaussian line (\textsc{phabs[apec + power + gaussian]}). Since the extended emission surrounding the nucleus may be of thermal origin, the extracted region may still contain thermal emission that needs to be accounted for. The fit is however largely unconstrained given the low number of counts. Instead, we keep the plasma temperature frozen to the value found in the previous section $kT=1.7$ keV. The abundance is also frozen to a 0.3 Solar value. We also require the Gaussian line energy to be frozen at 3.3 keV and the power law index to be frozen at 1.8 to obtain a reasonable fit. Even when freezing all of these parameters, the fit tends to favour a model in which the flux of the \textsc{apec} component is unrealistically small. Finally, we fit the spectrum of the nucleus with a thermal plasma model and absorbed power law with Gaussian line (model III). Similarly, we require that the temperature be frozen to $kT=1.7$ keV, the abundance to 0.3 Solar, the power law index to 1.8, the line energy to 3.3 keV and the dispersion to 300 keV (the average value from previous fits), to have the model converge. The results are shown in Table \ref{tab3}. Note that the cluster luminosity contributes to approximately one tenth of the total luminosity in the nuclear region.

It is important to note that PHABS model does not include the effects of Compton scattering, which can be important in heavily obscured circumnuclear media such as in IRAS F15307+3252. Therefore, we modified our model to include the \citet{Bri2011413} \textsc{torus} Monte Carlo calculations (model IV). This code self-consistently includes absorbed transmission, Compton scattering and fluorescence emission from a torus modelled as a conical section of a sphere with a smooth gas distribution with Solar abundance. The torus opening angle is a free parameter which effectively allows variable covering factors. Given the relatively low number of counts of our data, however, we fixed the opening angle and the inclination angles to median values of 60\,deg and 70\,deg, respectively, which is fairly typical for obscured AGN. We also require that the temperature of the thermal component be frozen at $kT=1.7$ keV, with an abundance of 0.3 Solar. We find a column density $N_{\rm H}$\,$>$\,2.5\,$\times$\,10$^{24}$\,cm$^{-2}$ and $L_{2-10~\rm{keV}}$\,$\sim$\,2\,$\times$\,10$^{45}$ erg s$^{-1}$. In this case, the thermal component accounts for less than 1 per cent of the $2-10$ keV emission. The correction for Compton scattering now implies a more realistic and higher intrinsic power than found with the \textsc{zphabs} in model III alone. Note that typical systematic uncertainties in determination of intrinsic luminosities span at least an order of magnitude \citep[][]{Gan2015449,Gan2016}, which means that the $2-10$ keV luminosity may be even larger. 

Finally, we attempt to model the emission by adding a complex absorber at the redshift of the source (\textsc{zpcfabs}) to the power law and Gaussian model, but the fit remained unconstrained even when freezing most parameters. We note that in all cases, there is a slight, soft excess at $0.3-0.6$ keV and that the rest-frame equivalent width of the Gaussian line varies between $\sim1-3$ keV, depending on the model used. 

We also repeat all of the fits using a different background region. We select a region located within an annulus of inner radius 1 arcsec and outer radius of 2.5 arcsec. The latter corresponds to the extended X-ray emission detected immediately surrounding the galaxy. This will allow us to subtract the majority of the thermal emission that contributes within the inner 1 arcsec. All results are shown in the bottom portion of Table \ref{tab3}. We fit the same models as in the previous case. We note that in this case, the complex absorber model and the model including the thermal component (model III) remained unconstrained. Model IV also required that we keep the power law index frozen to 1.8. Selecting a surrounding region of inner radius 1 arcsec and outer radius of 4.4 arcsec, encompassing all of the extended X-ray emission within $r<35$ kpc of the galaxy (Fig. \ref{fig3}), does not change our results.

\section{Discussion}

\subsection{Origin of the extended X-ray emission}

Using $XMM-Newton$, \citet{Iwa2005362} were the first to report evidence of X-ray emission associated with IRAS F15307+3252. They found the presence of a line-like feature in the observed $3-4$ keV range, as well as tentative evidence that the X-ray emission was extended beyond the PSF on scales of $r=10.5\pm$2.5 arcsec ($r=60-100$ kpc). Our $Chandra$ observations confirm the extended X-ray
emission surrounding IRAS F15307+3252. We find that this emission
extends at least to $r=35$ kpc (Fig. \ref{fig3}) and potentially out
to $r=160$ kpc (Fig. \ref{fig4}), in agreement with the results from
\citet{Iwa2005362}. 

However, using the ROSAT High Resolution Imager, \citet{Fab1996283} reported no evidence of X-ray emission associated with IRAS F15307+3252. They calculated a 3$\sigma$ upper limit of 3.5$\times10^{43}$ erg s$^{-1}$ in the $0.1-2.4$ keV range at the location of the galaxy. Our $Chandra$ observations imply that the total X-ray luminosity within $r=35$ kpc is $\sim3.8\times10^{43}$ erg s$^{-1}$ in the $0.1-2.4$ keV range, which appears to be inconsistent with the ROSAT results. \citet{Fab1996283} however identified an X-ray source located 13 arcsec to the south of the galaxy. They argued that this X-ray emission might be the core of a cluster of which IRAS F15307+3252 is a member. We have analysed the ROSAT images and find that several point sources are offset by several arcsecs compared to those seen in the $Chandra$ images. The point sources of our $Chandra$ images coincide however with several galaxies as seen in the optical Digitized Sky Survey (DSS). We therefore conclude that the ROSAT images were most likely misaligned, implying that emission seen in \citet{Fab1996283} and located 13 arcsec to the south is actually coincident with IRAS F15307+3252. 

Our $Chandra$ analysis showed as well that the extended emission is detected at a $>7\sigma$ level in the inner 20 kpc of the galaxy, and falls below a $2\sigma$ level beyond $r=35$ kpc (Fig. \ref{fig3}). Using the binned image in Fig. \ref{fig4}, we find that the large-scale extended X-ray emission is detected at a $3-6\sigma$ level at $r=30-100$ kpc, but only at a $2-3\sigma$ level at $r=100-160$ kpc. We confirm that this extended emission is soft, with no evidence of its existence in the hard X-ray images above a 2$\sigma$ level. The X-ray emission therefore clearly extends beyond the effective radius of the galaxy \citep[$r\sim12$ kpc in the $I$-band,][]{Far2002329}. Several studies have reported evidence that the galaxy is currently undergoing a merger or interaction \citep{Far2002329,Soi1994433,Liu1996470}. \citet{Iwa2005362} also highlighted the presence of a chain of galaxies located $\sim13$ arcsec (100 kpc) to the south of IRAS F15307+3252. This chain appears in the left panel of Fig. \ref{fig4}. Fig. \ref{fig2} further highlights the various knots and filaments within the giant elliptical itself. This could be a snapshot view of on-going groupings/mergers expected in a growing cluster environment. While the optical view shows rich sub-structure, none of this could be spatially resolved with $XMM-Newton$ in \citet{Iwa2005362} due to its relatively extended point-source response (PSF). The $Chandra$ X-ray image reveals that the X-ray emission is not only highly asymmetric, but that it also correlates with the eastern filament seen in the $HST$ image (Fig. \ref{fig3}), and extends in the southern direction to very large radii. The maximum extent of this X-ray emission seems to coincide with the chain of galaxies $\sim13$ arcsec to the south, but it is unclear if these galaxies belong to the cluster or are simply seen in projection. 

There are several possible origins for this extended X-ray emission. Because of its highly asymmetric morphology, it could be a highly beamed X-ray jet, originating from the AGN in IRAS F15307+3252. However, our VLA observations reveal no radio counterpart, making it remarkably radio faint. The 3$\sigma$ upper limit to the radio emission covering the extent of the X-ray emission out to $r=160$ kpc is $<$2.9 mJy at 1.4 GHz (using the AB0806 dataset) or $<$0.77 mJy at 1.4 GHz (using FIRST), and $<$0.24 mJy at 8.4 GHz (using the AB654 dataset) or $<$0.56 mJy at 8.4 GHz (using the deep AG0514 dataset). For all of these calculations, we calculate the upper limit defined as $3\times{\rm{rms}}\times\sqrt{area/beam}$. Fitting a power law model to the X-ray emission, we found in Section 3 that the emission, although very soft with $\Gamma=2.6\pm^{+0.6}_{-0.5}$, could be fit with a power law of flux density of 93$^{+27}_{-24}$ nJy at 1 keV. Its radio faintness could be explained by inverse Compton scattering of the cosmic microwave background (ICCMB), which predicts an increase in the X-ray to radio flux ratio with redshift as the CMB energy density increases \citep[$\big[{\nu}F_{\nu}{\big]}_x/\big[{\nu}F_{\nu}{\big]}_r$;][]{Sch2002569}. To be consistent with other studies on high-$z$ ICCMB X-ray jets \citep[e.g.][]{Che2004600}, at $z\sim1$ the $\big[{\nu}F_{\nu}{\big]}_x/\big[{\nu}F_{\nu}{\big]}_r$ ratio would need to be at most a couple dozen to 100 \citep[see also][]{Sim2016816}. Taking our radio upper limits, we find that the ratio at 1.4 GHz is at least 5000 (and even higher at 8.4 GHz). The jet-origin of the extended X-ray component is therefore implausible. 

Several authors have reported that some heavily absorbed AGN may scatter part of their light if a gap is present in the obscuration. The light may then ionize the surrounding regions even out to $\sim20$ kpc, as seen in the extended emission line region known as Hanny's Voorwerp \citep[e.g.][although a faded AGN could also explain this structure]{Lin2008389,Lin2009399}. Such a scenario may explain the extended, asymetric emission seen in IRAS F15307+3252, especially given the large polarisation fraction seen in spectropolarimetric observations \citep[$\sim18\%$;][]{Hin1995450}, which is suggestive of scattering in the narrow line region. We also find that the scattering fraction is $\sim$0.3 per cent when considering the $2-10$ keV emission of the plume ($\sim7\times10^{42}$ erg s$^{-1}$) compared to the intrinsic $2-10$ keV luminosity of the AGN ($\sim2\times10^{45}$ erg s$^{-1}$). This value is small, but agrees with other examples of scattering fractions \citep[][]{Tur1997488,Ued2007664,Win2009690}. However, the extremely large extent of the source ($r\sim160$ kpc) and the difference between the power law index of the extended emission compared to the AGN, make this scenario difficult to consider \citep[e.g.][]{Kee2015149}. Indeed, the extended emission can be described by a power law index of $2.6^{+0.6}_{-0.5}$, whereas the power law index of the intrinsic AGN is $1.7^{+0.6}_{-0.7}$. These two are inconsistent which each other at a 95.4$\%$ level. 

Wilman et al. (2003) and Alexander et al. (2005)\nocite{Wil2003338,Ale2005357} found evidence of soft X-ray emission associated with HyLIRGs and ULIRGs, and they suggested that its origin be thermal emission from star formation. However, the X-ray emission detected in these studies was significantly softer (kT$\sim0.3$ keV) and of lower luminosity ($\sim10^{42}$ erg s$^{-1}$) than that detected here. 

Based on $Spitzer$ observations, \citet{Rui2013549} obtained a rough estimate of the star formation rate in IRAS F15307+3252 ($3000\pm2000$ M$_\odot$ yr$^{-1}$). Given that the stellar mass of this galaxy is log$M_\star=11.36$ M$_\odot$ \citep{Com2013550}, the X-ray emission due to star formation processes such as X-ray binaries can therefore account for roughly $\sim50\%$ of the X-ray emission seen in IRAS F15307+3252 \citep{Leh2010724}. However, the $Chandra$ images reveal that the emission extends at least 3 times further ($r\sim35$kpc) than the effective radius of the galaxy ($r\sim12$ kpc). This makes the large extent of the X-ray emission difficult to explain if it were being driven by star formation. There is also no extended radio emission associated with IRAS F15307+3252 as would be expected if the vigorous star formation was fuelling radio emisison \citep{Bau2002124}. Our observations can however help constrain the star formation rate estimate for the galaxy. If all of the X-ray emission within 2 effective radii of the galaxy ($\sim24$ kpc) were due to star formation, then the star formation rate would be at most $\sim1800$ M$_\odot$ yr$^{-1}$ \citep{Leh2010724}. Similarly, our non detection of extended emission at 1.4 GHz implies that the star formation rate is at most $\sim2800$ M$_\odot$ yr$^{-1}$ based on the correlations from \citet{Bel2003586}. We stress however that the noise level for the 1.4 GHz data is quite high (see Table 2), and that our non detection at 8.4 GHz assuming a flat spectrum of $\alpha=0.5$ as typically seen in star forming galaxies, would bring down the star formation rate to at most $\sim1300$ M$_\odot$ yr$^{-1}$. Although this is still consistent with the value reported by \citet[][; $3000\pm2000$ M$_\odot$ yr$^{-1}$]{Rui2013549}, our calculations imply that the star formation rate may be on the lower bound of this estimate.

Based on the arguments mentioned above, we therefore argue that the extended X-ray emission surrounding IRAS F15307+3252, although highly asymmetric, is indicative of hot intracluster, or intragroup gas, being virialised in the potential of the galaxy. Fitting a thermal plasma model, we found in Section 3 that if we include all of the emission out to $r=160$ kpc, this gas has a temperature of $kT=2.2^{+1.7}_{-0.6}$ keV and a $0.3-8$ keV luminosity of $2.3^{+0.7}_{-0.6}\times10^{43}$ \ergps. The corresponding bolometric luminosity is $3.5^{+1.0}_{-0.9}\times10^{43}$ erg s$^{-1}$ in the $0.01-100$ keV range. Similarly, when only considering the emission within $r=35$ kpc, we find that $kT=1.7^{+1.7}_{-0.4}$ keV, the X-ray luminosity in the $0.3-8.0$ keV range is $1.5^{+0.5}_{-0.4}\times10^{43}$ \ergps with a bolometric $0.01-100$ keV luminosity of $2.4^{+0.9}_{-0.7}\times10^{43}$ \ergps. The cooling time that corresponds to this emission is 1.2$\times10^9$ Gyr within $r=35$ kpc, which is remarkably short. The corresponding entropy, defined as $S=kTn_e^{-2/3}$, is 23 keV cm$^2$. Such as small entropy is usually found in cool core clusters \citep[e.g.][]{Cav2009182,McD2013774}. 

The maximum mass deposition rate, in absence of heating, is given by
\begin{equation}
\dot{M}_{\rm I} = \frac{2{\mu}m_{\rm H}}{5k_{\rm B}}\frac{L}{T} \,  ,
\label{eq1}
\end{equation}
where ${\mu}m_{\rm H}$ is the mean gas mass per particle with $\mu=0.61$ and the luminosity is taken as the bolometric luminosity of the extended X-ray emission within $r<35$ kpc with $kT\sim1.7$ keV. Following this equation, if the gas were able to cool completely, a cooling flow of 57$M_{\rm{\odot}}$ yr $^{-1}$ would be produced. This is similar to nearby systems like Abell 2052 \citep[e.g.][]{Ode2008681}. 

Finally, we mention that with a star formation rate of approximately $3000\pm2000$ M$_\odot$ yr$^{-1}$ \citep{Rui2013549} and a stellar mass of log$M_\star=11.36$ M$_\odot$ \citep{Com2013550}, the specific star formation rate of the galaxy is on the order of $4-22$ Gyr$^{-1}$. This is significantly higher than the expected $z\sim1$ value based on the observed evolution of star formation in brightest cluster galaxies \citep{Web2015814,McD2016817}. If the star formation rate is indeed as high as reported by \citet{Rui2013549}, even when taking the lower bound as our data suggest, IRAS F15307+3252 might therefore represent a class of objects caught in a different evolutionary phase compared to the general population of clusters of galaxies and their brightest cluster galaxies. 

It is important to note, however, that the extended X-ray emission follows the expected temperature to X-ray luminosity correlation for groups and clusters. IRAS F15307+3252 falls directly at the intersection between these two systems \citep[e.g.][]{Pra2009498}. IRAS F15307+3252 might therefore represent a massive group (M$\sim10^{14}M_{\rm{\odot}}$ that will grow to become a Virgo-like cluster in the local Universe, increasing its mass and X-ray luminosity by a factor of a few over 7 Gyrs. We have therefore identified an object that is at a very interesting stage in its formation; likely not before long the transition to a cluster occurs.

\subsection{The hidden quasar}

We find that the nucleus spectrum of IRAS F15307+3252 is dominated by a line-like feature in the observed $3-4$ keV band. \citet{Iwa2005362} found a similar feature, but their $XMM-Newton$ observations could not resolve the AGN from the surrounding extended X-ray emission. This is now possible with the $Chandra$ observations. 

Table \ref{tab3} summarises the main models that were fit to the nucleus. Overall, we clearly detected the emission line located at $\sim3.3$ keV, corresponding to the redshifted 6.4 keV Fe K$\alpha$ line. This strongly indicates the presence of a reflection-dominated Compton-thick AGN in IRAS F15307+3252 (see for a review Fabian et al. 2000)\nocite{Fab2000112}. We stress that in model IV, we were able to account for this internal absorption using the \textsc{torus} model that accounts for absorbed transmission, Compton scattering and fluorescence emission. The resulting absorption is several $10^{24}$ cm$^{-2}$ and the total, unabsorbed non-thermal emission is $\sim2\times10^{45}$ erg s$^{-1}$ in the $2-10$ keV band (rest-frame). We note that the spectropolarimetry results from \citep{Hin1995450} also strongly suggest the presence of a hidden quasar in IRASF15307+3252. Applying a typical bolometric correction of 40-70 for quasars to this non-thermal emission \citep{Vas2007381,Elv199495}, we find that the bolometric luminosity of the AGN in IRAS F15307+3252 is roughly $0.8-1.4\times10^{47}$ erg s$^{-1}$.

The black hole mass was estimated by \citet{Iwa2005362} using the empirical relation based on the virial theorem for the broad line region and the linewidth of Mg$_{\rm II}$. They found $M_{\rm BH}\sim1.3\times10^9M_{\rm \odot}$. The bolometric luminosity of the AGN then corresponds to $50-90$ per cent of the Eddington limit. 

Finally, we note that the AGN is detected at radio wavelengths and that it is radio-loud with $L_{\rm 1.4 GHz}\sim3.5\times10^{25}$ W/Hz (Table \ref{tab2}). The spectral index for the AGN in the radio band was estimated to be $\alpha\sim1$ between 1.4 GHz and 8.4 GHz, while Fig. \ref{figsed} revealed the presence of a turnover at $\sim200$ MHz. Such a turnover is characteristic of a CSS source in IRAS F15307+3252. These sources are compact, powerful radio sources with typical scales of $r<15$ kpc. The radio images reveal that it is unresolved at $\sim0.7$ arcsec at 8.4 GHz, which corresponds to a scale of $\sim5$ kpc. \citet{Ode1998110} find a correlation between frequency turnover and linear projected size for CSS and gigahertz peaked-spectrum (GPS). For a turnover frequency of $\sim200$ MHz, the correlation predicts a size of 2-20 kpc, which is consistent with our target. We note that 3C 186 is another example of a powerful CSS quasar embedded in a cluster at $z\sim1$ \citep[e.g.][]{Sie2010722}, consistent with it being a young radio source in its early stages of evolution. This may also be the case for IRAS F15307+3252. Finally, we note that the NVSS integrated flux (8.1$\pm0.5$ mJy) is slightly larger than the flux reported in the FIRST survey ($5.9$mJy). This could be due to intrinsic variations in the quasar flux.

\subsection{Implications for cluster surveys}
HyLIRGs are thought to play an important role in the cosmic cycle of
galaxy evolution \citep{Hop2006163}. They could be the most powerful
manifestation of galaxy mergers, in a phase where gas is flowing
towards the centre of the galaxy, enhancing star formation and nuclear
activity. They are therefore thought to play a key role in explaining
the high redshift origin of massive elliptical and S0 galaxies
\citep{Fra1994427}; an important fraction of stars in present day
galaxies would have been formed during these evolutionary
phases. HyLIRGs also probably represent the pinnacle of star formation
activity, caught at the most vigorous stage of galaxy formation in
which the most massive galaxies and their black holes are assembling. Located
at intermediate redshifts, they are comparatively nearby compared to
their distant cousins, the sub-mm galaxies. They are therefore unique
laboratories for studying extremely high star formation and
its connection to black hole and galaxy growth at a critical stage, likely
not long before the switch-over to low Eddington growth. IRAS F15307+3252 is no exception to this. 

In terms of X-ray properties, the most complete X-ray study of HyLIRGs to date analysed only 14 objects \citep{Rui2007471}. Although previous studies indicated a predominance of the starburst over the AGN phenomenon
(Franceschini et al. 2003), \citet{Rui2007471} found that all the X-ray detected HyLIRGs in their sample had AGN-dominated X-ray spectra, five of which were Compton
thick. Most importantly, they found
that the hard X-ray to FIR luminosity ratio remained constant with redshift, suggesting that
the AGN and starburst phenomena are physically connected in HyLIRGs, and that beyond $z=0.5$, the
majority of HyLIRGs have $L_{\rm 2-10 keV}>0.01L_{\rm{FIR}}$. When considering our $L_{\rm 2-10 keV}$ estimate of $\sim2\times10^{45}$ erg
s$^{-1}$, we find that the ratio for IRAS F15307+3252 is $L_{\rm 2-10
  keV}\sim0.01L_{\rm{FIR}}$, in agreement with
\citet{Rui2007471}. 

Despite the small number of HyLIRGs analysed in the
X-ray band, \citet{Rui2007471} also found that HyLIRGs are a heterogeneous population with, perhaps, two
or three classes: one with broad-band properties similar to local QSOs, another showing additional starburst
components but no merger signatures, and finally those showing both starburst and AGN components. Even
more striking is the emerging trend of results showing that the less luminous cousins of HyLIRGs (ULIRGs) appear to be intrinsically X-ray fainter than typical Seyfert galaxies (e.g.
Teng et al. 2014; see also the recently revised X-ray to mid-IR relation by Stern 2015 for powerful quasars)\nocite{Ste2015807,Ten2014}. If HyLIRGs were also intrinsically X-ray faint,
they would be offset from the X-ray to mid-IR relations (Gandhi et al. 2009; Asmus et al. 2015; Stern 2015)\nocite{Gan2009502,Ste2015807,Asm2015454}. Using WISE data, we find that IRAS F15307+3252 has a rest-frame IR luminosity at 12 $\mu$m of $\sim3.1\times10^{46}$ erg s$^{-1}$ and at 6 $\mu$m of $\sim1.2\times10^{46}$ erg s$^{-1}$. This places the galaxy at the intersection between the Gandhi et al. 2009 and Stern 2015 X-ray to mid-IR relations. The characteristics of the intrinsic HyLIRG population (especially in X-rays) are therefore far from clear, and
increasing the sample is needed. 

Using targeted $Chandra$ observations, we have found evidence for 
extended X-ray emission surrounding IRAS F15307+3252, consistent with a
massive group in the process of transforming into a Virgo-like
cluster. IRAS 09104+4109, also known as MACS J0913.7+4056, is another example of a HyLIRG located at the center of a cluster \citep[$z=0.442$;][]{Kle1988328,Hla2012421,OSu2012424}. Similarly to IRAS F15307+3252, IRAS 09104+4109 hosts a heavily obscured Type 2 AGN \citep[$N_{\rm H}=1-5\times10^{23}$ cm$^{-2}$;][]{Vig2011416,Iwa2001321,Hla2013431}. This is also the case for the massive Phoenix cluster ($z=0.596$) which hosts a ULIRG at its center with a heavily obscured AGN \citep[$N_{\rm H}\sim3-5\times10^{23}$ cm$^{ 2}$;][]{Ued2013778,Toz2015580}. 

Finally, we mention that the eROSITA all-sky survey will not be sensitive to faint extended sources such as IRAS F15307+3252 ($\sim10^{-14}$ erg cm$^{-2}$ s$^{-1}$) with a flux limit of $\sim3.4\times10^{-14}$ erg cm$^{-2}$ s$^{-1}$ \citep{Mer2012}. This sky survey will therefore miss an important population of still-forming groups and/or clusters similar to IRAS F15307+3252. The Wide Field Imager, onboard the \textsc{athena} X-ray observatory, will however easily detect the extended X-ray emission surrounding IRAS F15307+3252. Assuming a fixed temperature of 2.2 keV, an abundance of 0.3 and an observed flux of $4.7\times10^{-15}$ erg s$^{-1}$ cm$^{-2}$, we simulated the observations using the ARF and RMF instrument response files of \textsc{athena}\footnote{Note that we have used the March 2015 response matrices and background files, http://www.mpe.mpg.de/ATHENA$-$WFI/ public/resources} and the \textsc{fakepha} command in \textsc{sherpa}. We consider the instrumental and diffuse background for an extended sources. For an on-axis observation, we obtain roughly 4000 counts for a 100 ks exposure, which would be sufficient to analyse in detail the properties of the hot X-ray gas.

In terms of large mm-wave surveys, the number of known high-redshift galaxy clusters has increased dramatically in recent years, largely due to the success of these surveys in using the Sunyaev-Zel'dovich (SZ) effect to select massive clusters at all redshifts. These SZ surveys include the South
Pole Telescope (SPT) (Staniszewski et al. 2009; Vanderlinde et al. 2010;
Reichardt et al. 2013; Bleem et al. 2014) and the Planck satellite
(Planck Collaboration 2011, Planck-29 2013). However, none of these
have the sensitivity to detect $\sim10^{14}M_\odot$ groups at
$z\sim1$, although the upcoming SPT-3G might be able to reach this
limit \citep{Ben2014}. 

In summary, targeted follow-up X-ray observations of similar HyLIRGS to
IRAS F15307+3252 are needed before understanding the general
properties of these objects. We stress once more that if the eROSITA all-sky
survey will be missing such objects, this means that they will be
missing an important population of still-assembling groups and
clusters of galaxies. This will have important cosmological
implications for our understanding of how groups and clusters form.

\section{Conclusions}

IRAS F15307+3252 is a classical HyLIRG located at z=0.93, hosting a radio-loud AGN with $L_{\rm 1.4 GHz}\sim3.5\times10^{25}$ W/Hz. Based on the
new $Chandra$ observations obtained, we have identified extended, highly
asymmetric emission to the south. This emission has a temperature of $\sim2$ keV and no radio counterpart, although the radio observations reveal that the galaxy hosts a CSS. We therefore argue that the extended X-ray emission
is of thermal origin with a temperature ($\sim2$ keV) and bolometric
X-ray luminosity ($\sim3\times10^{43}$ erg s$^{-1}$) that agree with
the expected $L_{\rm X-ray}-T$ correlation for groups and clusters of
galaxies. The relatively small X-ray luminosity implies that IRAS
F15307+3252 is more consistent with a massive group that is forming,
instead of a cluster. By $z\sim0$ it should reach a mass that is
similar to the Virgo cluster. We have therefore identified an object
which is at a particularly interesting phase in its formation, transiting
between from a group to a cluster of galaxies. We also find that the extended gas has a
remarkably short cooling time of $1.2$ Gyrs and a central entropy of 23
keV cm$^{-2}$. Finally, we note that the spectrum of the nucleus is
dominated by a redshifted 6.4 keV Fe K$\alpha$ emission line, in
agreement with previous studies. Fitting spectral model to the nuclear
emission, we find that its unabsorbed $2-10$ keV luminosity is at least a few $10^{45}$ erg
s$^{-1}$ and that the bolometric luminosity of the AGN corresponds to at least 50 per cent of the Eddington limit for the black hole. 

\section*{Acknowledgments}

JHL is supported by NSERC through the discovery grant and Canada Research Chair programs, as well as FRQNT. KI acknowledges support by the Spanish MINECO under grant AYA2013-47447-C3-2-P and MDM-2014-0369 of ICCUB (Unidad de Excelencia 'Mar\'ia de Maeztu'). P.G. is supported by STFC (grant reference 
ST/J003697/2) and ACF is supported by ERC Advanced grant 340442. Some of the data used in this paper were obtained from the Mikulski Archive for Space Telescopes (MAST). STScI is operated by the Association of Universities for Research in Astronomy, Inc., under NASA contract NAS5-26555. Support for MAST for non-HST data is provided by the NASA Office of Space Science via grant NNX09AF08G and by other grants and contracts. We also thank the referee for his useful comments, as well as Roberto Antonucci for comments on the polarization of the target. Finally, we thank Anne Lohfink for helpful discussions about ROSAT and Tracy Clarke for her help with VLITE.

\bibliographystyle{mn2e}
\bibliography{bibli}
\end{document}